\documentclass[twocolumn,twoside,slac_two]{revtex4}
\usepackage{graphicx}
\usepackage{fancyhdr}
\usepackage{graphicx}
\usepackage{dcolumn}
\usepackage{amsmath}
\usepackage{epsfig}
\usepackage{multirow}
\RequirePackage{xspace}
\usepackage{relsize}

\def\barpk{{\raise.35ex\hbox
{${\sss (}$}}--{\raise.35ex\hbox{${\sss )}$}}}
\def\bbarp{\hbox{$B$\kern-0.9em\raise1.4ex\hbox{\barpk}}}
\def\beq{\begin{equation}}
\def\eeq{\end{equation}}
\def\bea{\begin{eqnarray}}
\def\eea{\end{eqnarray}}
\def\nn{\nonumber}
\def\sss{\scriptscriptstyle}
\def\roughly#1{\mathrel{\raise.3ex\hbox
{$#1$\kern-.75em\lower1ex\hbox{$\sim$}}}}

\def\bd{B_d^0}
\def\bs{B_s}
\def\bdbar{{\bar B}_d^0}
\def\btod{{\bar b} \to {\bar d}}
\def\btos{{\bar b} \to {\bar s}}

\def\fT{f_{\sss T}}
\def\fL{f_{\sss L}}
\def\fTfL{f_{\sss T}/f_{\sss L}}
\def \Kbar{\bar K}

\def\Abar{{\bar A}}

\def\Abar {\kern 0.18em\overline{\kern -0.18em A}{}\xspace}
\def\Kbar {\kern 0.18em\overline{\kern -0.18em K}{}\xspace}
\def\Bbar {\kern 0.18em\overline{\kern -0.18em B}{}\xspace}
\def\Dbar {\kern 0.18em\overline{\kern -0.18em D}{}\xspace}

\def\barpk{{\raise.35ex\hbox
{${\sss (}$}}--{\raise.35ex\hbox{${\sss )}$}}}
\def\bbarp{\hbox{$B$\kern-0.9em\raise1.4ex\hbox{\barpk}}}
\def\beq{\begin{equation}}
\def\eeq{\end{equation}}
\def\bea{\begin{eqnarray}}
\def\eea{\end{eqnarray}}
\def\nn{\nonumber}
\def\sss{\scriptscriptstyle}
\def\roughly#1{\mathrel{\raise.3ex\hbox
{$#1$\kern-.75em\lower1ex\hbox{$\sim$}}}}

\def\bd{B_d^0}
\def\bs{B_s}
\def\bdbar{{\bar B}_d^0}
\def\btod{{\bar b} \to {\bar d}}
\def\btos{{\bar b} \to {\bar s}}

\def\fT{f_{\sss T}}
\def\fL{f_{\sss L}}
\def\fTfL{f_{\sss T}/f_{\sss L}}
\def \Kbar{\bar K}

\def\Abar{{\bar A}}

\begin{document}


\title{Polarization Studies: testing explanations of the $B\to\phi
K^*$ puzzle and $B \to VT$ decays}

%

\author{Alejandro Szynkman}
\affiliation{
Physique des Particules, Universit\'e de
Montr\'eal,C.P. 6128, succ. centre-ville, Montr\'eal, QC,
Canada H3C 3J7}

\begin{abstract}
It has been observed in $B\to\phi K^*$ and $B^{0(+)}\to \rho^{0(+)}
K^{*0}$  that the fraction of transverse decays, $\fT$, and the
fraction of longitudinal decays, $\fL$, are roughly equal, in
opposition to the naive expectation, $\fTfL \ll 1$. If one requires a
single explanation of all polarization puzzles, two possibilities
remain within the standard model (SM): penguin annihilation and
rescattering. We examine the predictions of these two explanations for
$\fTfL$ in $\btod$ decays. We also study polarization observables in
$B \to VT$ decays ($V$ is a vector and $T$ is a tensor meson) to probe
whether the two SM explanations account for the $\fTfL$ ratio in this
type of decays and to further investigate the two new-physics
scenarios which explain the data in $B\to\pi K$ and the $\phi( \rho)
K^*$ polarization measurements.
\end{abstract}

\maketitle

\thispagestyle{fancy}


\section{Introduction}

In certain $B\to V_1V_2$ decays ($V_i$ is a light charmless vector
meson) dominated by $\btos$ penguin transitions in the standard model
(SM), an unexpected observation has been made. Considering that the
final particles have spin 1, it is straightforward to see that a given
$B$ decay into two vector mesons can be analyzed as three separated
$B$ decays -- one for each polarization of the vector states (one
longitudinal, two transverse) -- into two spinless particles. Naively,
the transverse amplitudes are suppressed by a factor of size $m_{\sss
V}/m_{\sss B}$ ($V$ is one of the vector mesons) with respect to the
longitudinal amplitude. As such, one expects the fraction of
transverse decays, $\fT$, to be much less than the fraction of
longitudinal decays, $\fL$. However, it has been measured that these
two fractions are roughly equal in $B\to\phi K^*$~\cite{phiK*} and
$B^{0(+)}\to \rho^{0(+)} K^{*0}$~\cite{belle:rhokst,babar:rhokst} (see
Table~\ref{table1} for recent measurements).

\begin{table}[tbh]
\center
{
\small
\begin{tabular}{lccccc}
\hline
\vspace{-0.3cm}\\
Mode & ${\cal B}(10^{-6})$ & $f_{\sss L}$ & $f_\perp$ \\
\vspace{-0.3cm}\\
\hline\hline
\vspace{-0.3cm}\\
$\phi K^{*0}$ \cite{phiK*0, belle:phikst, cdf:phikst} &
  ${ 9.5}\pm{0.9}$ & ${ 0.49}\pm{0.04}$ & ${ 0.27}^{+0.04}_{-0.03}$ \\
$\phi K^{*+}$ \cite{phiK*, belle:phikst, babar:phikstpl} &
  ${ 10.0}\pm 1.1$ & ${ 0.50}\pm 0.05$ & ${0.20}\pm 0.05$ \\
$\rho^+ K^{*0}$ \cite{belle:rhokst, babar:rhokst} &
${ 9.2}\pm 1.5$ &  ${ 0.48}\pm 0.08$ \\
$\rho^0 K^{*0}$ \cite{babar:rhokst} &
${ 5.6}\pm{1.6}$  & ${ 0.57}\pm0.12$ \\
$\rho^- K^{*+}$ \cite{babar:rhokst} &
({$< 12.0$}) \\
$\rho^0 K^{*+}$ \cite{babar:rhokst} &
({$3.6^{+1.9}_{-1.8}$}) & $(0.9\pm 0.2)$ \\
{$\phi K_2^{*}(1430)^0$} \cite{phiK*0} & ${7.8}\pm{1.3}$ & ${0.85}\pm 0.08$  & ${0.05}\pm0.05$\\
\vspace{-0.3cm}\\
\hline\hline
\end{tabular}
}
\caption{Measurements of the branching fraction ${\cal B}$,
longitudinal polarization fraction $\fL$ and fraction of parity-odd
transverse amplitude $f_\perp$, for $B\to\phi K^*$, $\rho K^*$, and
$\phi K_2^{*}(1430)^0$, expected to proceed through a $\btos$
transition~\cite{pdg, hfag}. Numbers in parentheses indicate
observables measured with less than 4$\sigma$ significance.}
\label{table1}
\end{table}

The differences between the measurements and the naive expectations
could be interpreted in favour of the presence of physics beyond the
SM~\cite{NP}, though none of these discrepancies has been
statistically significant. On the other hand, two explanations remain
as possible solutions within the SM when they are considered one at a
time: penguin annihilation~\cite{Kagan} and
rescattering~\cite{soni,SCET}.

These two explanations account for a large $\fTfL$ in $\btos$
decays. However, the key point is that a large $\fTfL$ is also
predicted in certain $\btod$ decays~\cite{DMV}. The measurement of
$\fTfL$ in these $\btod$ decays will allow us to test penguin
annihilation and rescattering as the explanations of the observed
$\fTfL$ ratio in $B\to\phi K^*$ decays~\cite{res}. Besides, we also
investigate their predictions in $B \to VT$ decays ($T$ is a tensor
meson)~\cite{vt}. Since in this case there are too three
polarizations,  $\fTfL$ can be measured. This ratio has been
experimentally determined in $B\to\phi K_2^*$~\cite{phiK*0}  (see
Table~\ref{table1}) and it is small. The potential solutions must also
explain $\fTfL$ in $B \to V T$ decays. In this paper, we study this
matter, both within the SM and assuming new physics (NP).

\section{SM Explanations of $\fTfL$ in $B\to\phi K^*$} 

We focus on $B\to V_1V_2$ decays. In this case the amplitude for the
process is given in the linear polarization basis by
\beq
M = A_0 \varepsilon_1^{*\sss L} \cdot \varepsilon_2^{*\sss L} 
- {1 \over \sqrt{2}} A_\| {\vec\varepsilon}_1^{*\sss T} \cdot
  {\vec\varepsilon}_2^{*\sss T}
- {i \over \sqrt{2}} A_\perp {\vec\varepsilon}_1^{*\sss T} \times
  {\vec\varepsilon}_2^{*\sss T} \cdot {\hat p} ~, 
\label{amp}
\eeq
where the polarizations of the final-state vector mesons
($\varepsilon_i^*$) are either longitudinal ($A_0$), or transverse to
their directions of motion and parallel ($A_\|$) or perpendicular
($A_\perp$) to one another. Along this article we also make use of
the helicity basis for the transverse polarizations, where $A_\pm=(A_\|
\pm A_\perp)/\sqrt{2}$.

The relative fractions into $V$ meson states with longitudinal and
transverse polarizations is
{\small
\beq
\fL = {|A_0|^2 \over |A_0|^2 + |A_+|^2 + |A_-|^2}
, \fT = {|A_+|^2 + |A_-|^2 \over |A_0|^2 + |A_+|^2 + |A_-|^2}, \nonumber
\eeq}
\hspace*{-0.2cm} where $f_{\sss T}=(1-f_{\sss L})$. Moreover, we can
define the relative fraction into final states with perpendicular
polarizations as
{\small
\beq f_\perp = {|A_\perp|^2 \over |A_0|^2 + |A_+|^2 + |A_-|^2}
~~.~~~~ \nonumber 
\eeq}
\hspace*{-0.27cm} As we noted in the introduction, final-state
particles in $B \to VT$ decays have also three possible
polarizations. Therefore, the amplitude for these decays can be
decomposed in the same way as in Eq.~(\ref{amp}).

We stated above that there are two potential SM solutions for the
polarization  puzzle observed in $B\to\phi K^*$. In this section
we review these two explanations.

We begin with penguin annihilation in the context of QCD factorization
(QCDF)~\cite{BBNS}. Normally, annihilation contributions are expected
to be small as they are higher order in the $1/m_b$ expansion, and
thus ignored.  However, within QCDf, it is possible that the
coefficients of these terms are large~\cite{Kagan}. Penguin
annihilation is not calculable in QCDf because of divergences which can
be parametrized in terms of unknown quantities chosen to fit the data
in $B\to\phi K^*$. The penguin annihilation amplitude arises only
from penguin diagrams with an internal $t$ quark.

We now discuss the second explanation:
rescattering~\cite{soni,SCET}. Rescattering effects involving charm
intermediate states can generate large transverse polarization. In a
particular picture~\cite{soni} of this solution, heavy charm mesons
rescatter to $B\to\phi K^*$. Then, if the transverse polarization is
not reduced in the scattering process, this mechanism will lead to a
large $\fTfL$.

With the previous paragraph we complete our brief review about the different
physical origin of the two explanations. We now take a look at the
similarities of calculation. In order to see this, consider the
penguin contribution ${\cal P}\!_q$ for the decay ${\bar b} \to {\bar
q} q' {\bar q}'$ ($q=d,s$, $q'=u,d,s$):
\begin{eqnarray}
\label{penguinamp}
{\cal P}\!_q & = & V_{ub}^* V_{uq} P_u + V_{cb}^* V_{cq} P_c +
V_{tb}^* V_{tq} P_t \nn\\ & = & V_{cb}^* V_{cq} (P_c - P_u) + V_{tb}^*
V_{tq} (P_t - P_u)~. \nn
\end{eqnarray}
Both in penguin annihilation and rescattering, the effect of the
dominant contribution to the transverse amplitudes is simply the
addition of one term in $P_t$ and $P_c$ respectively. Below we
follow the following prescription: we take into account the additional SM
effects by adding a single amplitude to represent the dominant
contribution to the transverse amplitudes.

\section{$B \to \rho\rho$ Decays}

Both penguin annihilation and rescattering explain the $\fTfL$ ratio
in the $\btos$ decay $B\to\phi K^*$ by modifying the penguin
amplitude. A similar modification must appear in some $\btod$
decays. Then, the question that we intend to reply in this section is
how their effects appear in $B \to \rho\rho$ decays.

Within the diagrammatic approach \cite{GHLR}, the three $B \to
\rho\rho$ amplitudes are given mainly by three diagrams: the
color-favored and color-suppressed tree amplitudes $Tr$ and $C$, and
the gluonic penguin amplitude $P$,
\bea
-\sqrt{2} A(B^+ \to \rho^+ \rho^0) & = & Tr + C ~, \nn\\
- A(\bd \to \rho^+ \rho^-) & = & Tr + P + R ~, \nn\\
-\sqrt{2} A(\bd \to \rho^0 \rho^0) & = & C - P - R ~. \nn
\eea
Here $R$ stands for the single extra term arising from the new penguin
annihilation or rescattering contribution. Some immediate conclusions
can be extracted from the above equations. First, since a modification
of $P$ is involved, $\fTfL$ in $B^+ \to \rho^+ \rho^0$ will not be
affected. This agrees with observation (see
Table~\ref{table2}). Second, in order to calculate $\fTfL$, it is
necessary to estimate the size of $R$. As discussed earlier,
rescattering and penguin annihilation affects the penguin amplitude
$P$; thus, $|R| \sim |P|$. This shows that $\fTfL$ is expected to be
small in $\bd \to \rho^+ \rho^-$, since it is proportional to
$|R|^2/|Tr|^2 \sim |P|^2/|Tr|^2$. This also agrees with observation
(Table~\ref{table2}). Finally, $\fTfL$ can be large in $\bd \to \rho^0
\rho^0$ since the contributions to the transverse and longitudinal
polarizations are the same size. It will be interesting to measure
this precisely.

\begin{table}[tbh]
\center
\begin{tabular}{lcccccc}
\hline
\vspace{-0.3cm}\\
Mode & ${\cal B}(10^{-6})$ & $f_{\sss L}$ \\
\vspace{-0.3cm}\\
\hline\hline
\vspace{-0.3cm}\\
$\rho^0\rho^+$ \cite{rho+rho0} &
  $18.2 \pm 3.0$  & ${0.912}^{+0.044}_{-0.045}$
 \\
$\rho^+\rho^-$ \cite{rho+rho-} &
  ${24.2}^{+3.0}_{-3.2}$ & $0.976^{+0.028}_{-0.024}$ 
 \\
$\rho^0\rho^0$ \cite{rho0rho0} &
  $(1.07 \pm 0.38)$  & $({0.86}^{+0.12}_{-0.14})$ 
\\
\hline\hline
\end{tabular}
\caption{Measurements of the branching fraction ${\cal B}$ and
longitudinal polarization fraction $\fL$ for $B^+$ and $B^0_d$ meson
decays into $\rho \rho$ final states. Numbers in
parentheses indicate observables measured with less than 4$\sigma$
significance.}
\label{table2}
\end{table}

There are some further tests to be performed. Since there is only one
added amplitude, one has $|A_+(\bd \to \rho^0 \rho^0)| =
|\Abar_-(\bdbar \to \rho^0 \rho^0)|$, and similarly for $A_-$ and
$\Abar_+$ (where $A_{\pm}$ are given in the helicity basis and
$\Abar_{\pm}$ stands for the corresponding amplitudes in the
$CP$-conjugated decay).  If this is not found, penguin annihilation
and rescattering will be ruled out. Another way to probe the SM
explanations is by using SU(3) since the extra transverse amplitudes,
$|R|$ ($\btod$) and $|R'|$ ($\btos$), are related by flavor
symmetries. This allows us to estimate $\fTfL$ in $\bd \to \rho^0
\rho^0$ from $B^+ \to \rho^+ K^{*0}$ decays. If we assume that SU(3)
is an exact symmetry, we obtain an explicit relation between $|R|$ and
$|R'|$: $R = |V_{td(cd)}/V_{ts(cs)}| R'$ in penguin annihilation
(rescattering). Thus, we can obtain the $\fTfL$ ratio in  $\bd \to
\rho^0 \rho^0$ by using experimental data on $B^+ \to \rho^+ K^{*0}$
decays, and we find
\bea
|A_{\sss T} (B^+ \to \rho^+ K^{*0})|^2 & = & (5.10 \pm 1.14) \times
 10^{-16}~{\rm GeV}^2 ~, \nn\\
|A_{\sss L} (\bd \to \rho^0 \rho^0)|^2 & = & (2.10 \pm 0.81) \times
 10^{-16}~{\rm GeV}^2 ~, \nn
\eea
leading to
\beq
\fTfL (\bd \to \rho^0 \rho^0) = |V_{td}/V_{ts}|^2 \, (2.43 \pm 1.08) ~.\nn
\eeq
This result agrees with data taken directly from $\bd \to \rho^0 \rho^0$ (see
Table~\ref{table2}):
\beq
\fTfL (\bd \to \rho^0 \rho^0) = (1-\fL)/\fL = 0.16 \pm 0.15 ~. \nn
\eeq
The agreement is good because of the large errors, making the
hypothesis of no violation of SU(3) a consistent assumption at this
accuracy level.  Equally, the measurement does not give a definite
answer as to whether $\fTfL$ is large or small. Another point is related to
this: if central values are taken, $\fTfL$ is not large after
all. This shows that $\fTfL$ is {\it not} guaranteed to be large in
$\bd \to \rho^0 \rho^0$. The reason for this is that, due to the
additional amplitude $C$, $\fL$ can be big, making $\fTfL$
small. There is a further complication: if $C$ contributes
significantly to the transverse polarization, as it could be the case
in QCDf, then a test of the explanations might be performed by means
of a time-dependent angular analysis (see~\cite{res} for details). The
lesson here is that it is best to consider $\btod$ decays for which
$\fTfL$ is expected to be large and which receive only one dominant
contribution to the transverse polarization.

\section{U-spin Pairs}

We have previously stressed the idea of measuring $\fTfL$ in $\btod$
decays. But this raises the question: how do we choose the $\btod$
decay to study? U-spin symmetry can help us to investigate this
issue. Pairs of $B$ decays which are related by U-spin are given in
Ref.~\cite{Gronau}. In $B\to VV$ form, these are
\begin{enumerate}
\item $\bd \to K^{*+}\rho^-$ and $\bs \to \rho^+ K^{*-}$~,
\item $\bs \to K^{*+} K^{*-}$ and $\bd \to \rho^+\rho^-$~,
\item $\bd \to K^{*0}\rho^0$ and $\bs \to \Kbar^{*0}\rho^0$~,
\item $B^+ \to K^{*0} \rho^+$ and $B^+ \to \Kbar^{*0} K^{*+}$~,
\item $\bs \to K^{*0} \Kbar^{*0}$ and $\bd \to \Kbar^{*0} K^{*0}$~.
\end{enumerate}
In all cases, the first decay is $\Delta S = 1$ ($\btos$); the second
is $\Delta S = 0$ ($\btod$).  The procedure here is to measure the
polarizations in the $\btos$ decay, and compare them with the
measurements in the corresponding $\btod$ decay.

As noted in the past section, the best $\btod$ decays to be considered
in the tests are those with an expected large $\fTfL$ ratio and with
only one contribution to the transverse amplitude. Given this, the best
possibilities are the last two pairs: (i) $B^+ \to K^{*0} \rho^+$
($\btos$) and $B^+ \to \Kbar^{*0} K^{*+}$ ($\btod$) and (ii) $\bs \to
K^{*0} \Kbar^{*0}$ ($\btos$) and $\bd \to \Kbar^{*0} K^{*0}$
($\btod$). We urge the measurement of $\fTfL$ in these pairs of decays.

The explanations of $\fTfL$ in $B\to\phi K^*$ then make three
predictions :
\begin{itemize}

\item $\fTfL$ is expected to be large in both the $\btos$ decay and
the corresponding $\btod$ decay.

\item $|A_{+}|$ and $|{\bar A}_{-}|$ are expected to be equal in both the
$B$ and ${\bar B}$ decays, and similarly for $A_{-}$ and
${\bar A}_{+}$.

\item $R'$ and $R$ can be extracted from the $\btos$ and
$\btod$ decays, respectively. These should be related by SU(3)
(including SU(3) breaking).

\end{itemize}
If any of these predictions fail, penguin annihilation and
rescattering are ruled out in the U-spin limit or for small U-spin
breaking.

Since the ratio of $\fTfL$ in these pairs of decays measures SU(3)
breaking, an additional test can be made. If one ignores SU(3) breaking, 
one has the following prediction
\beq
({\fTfL})_{\btod} = ({\fTfL})_{\btos} ~. 
\label{test}
\eeq
The breaking of SU(3) in the above equation is model dependent. If it
is found experimentally that the above relation is broken badly, then
particular models of penguin annihilation and rescattering will have
to invent a mechanism to generate large SU(3)-breaking effects or they
will be ruled out. In other words, Eq.~\ref{test} can be used to
constrain those specific models.

Up to here, we have not distinguished penguin annihilation and
rescattering since their effects are very similar. We will see now
that there is a possible way to differentiate them. As noted above,
rescattering involves only a change to $P_c$, while penguin
annihilation involves only $P_t$. However, the weak phase of these
pieces in $\btod$ decays is different: $\phi$(rescattering)$\sim 0$,
$\phi$(penguin annihilation)$\sim -\beta$. If this weak phase can be
measured, one could distinguish penguin annihilation and
rescattering. The measurement of that phase can be made by performing
a time-dependent angular analysis of the $\btod$ decay.  One has to
focus on observables which provide information about the relative phase
of the transverse amplitudes in the direct and the $CP$-conjugated
decays (see~\cite{res} for a more detailed discussion).  Of the decays
pointed out as those which satisfy the necessary requirements to carry
out meaningful tests, there is only one for which a time-dependent
angular analysis can be done: $\bd \to \Kbar^{*0} K^{*0}$.

\section{$B \to V T$ Decays}

So far we have been discussing $B \to V V$ decays. In the current
section we concentrate on $B \to V T$ processes. As mentioned in the
introduction, these decays are also analyzed in terms of three
polarizations and the $\fTfL$ ratio can be measured. 

The various explanations must account for the $\fTfL$ data in both
$B\to V_1V_2$ and $B \to VT$ decays. We examine this question now,  both
in the SM and with NP.

We first briefly review the SM (naive) prediction for $\fTfL$. In the
large-energy effective theory~\cite{formfactors}, it can be shown that
the $B \to T$ form factors are expressible in terms of two universal
quantities. Depending on the relative magnitude of these quantities,
three different scenarios are possible. It is found that two of them
predict $\fTfL \ll 1$, whereas the other $\fTfL \sim 1$. However,
since the last case is in contradiction with the experimental results
for $B \to \phi K_2^*$ (see Table~\ref{table1}), we conclude that the
SM naively predicts $\fTfL \ll 1$ in $B\to VT$ decays.
 
The next question is:  which are the penguin annihilation and
rescattering predictions for $\fTfL$? In order to answer this
question, we must establish whether or not the individual explanations
depend on the final-state particles. If they do not, then the
prediction for $\fTfL$ in $B\to\phi K_2^*$ will be the same as that in
$B\to\phi K^*$, which is in disagreement with experiment. The
calculation of penguin annihilation does depend on the final-state
wave function. Thus, it is possible that $\fTfL$ is small in $B\to\phi
K_2^*$ for the three cases discussed above, in agreement with
experiment. Within the rescattering solution, it is again possible to
choose parameters in order to obtain a  small $\fTfL$ in $B\to\phi
K_2^*$. These arguments can be extended to any other $B \to V T$
decay. Therefore, since there is a new set of parameters for each
final state, and it is virtually impossible to calculate the values of
the parameters, we conclude that both penguin annihilation and
rescattering are viable, but not very convincing.
 
Since the CP measurements in many penguin decays that proceed through
$\btos$ transitions~\cite{piK} and the polarization measurements in
some $B\to V_1V_2$ ($\btos$) appear to be in conflict with naive SM
expectations decays, it is not unreasonable to attempt to
understand the data assuming new physics. The important question to
ask is then the following: can we find a unified new-physics
explanation for all the discrepancies so far reported in measurements
of pure-penguin or penguin-dominated decays? And, what would we expect
for $B \to V T$ decays if NP is assumed? After considering a general
parametrization of NP as in Ref.~\cite{NP}, it appears that the NP
scenario is the same as that of the SM -- the prediction of $\fTfL$ in
$B \to \phi K_2^*$ depends on the values of unknown parameters and
with the proper choice it can be made consistent with the experimental
results (see~\cite{vt} for details). However, the difference is that,
with penguin annihilation and rescattering, the parameters are
essentially incalculable, while the NP prediction depends on form
factors.  Although the values of these form factors are not very well
known at the moment, they can be calculated. We strongly urge that the
$B\to T$ form factors be computed. In other decays like $B \to \rho
K_2^*$, the prediction for $\fTfL$ is the same as that of the SM since
the NP does not affect these processes. It will be important to
measure the polarization particularly in  $\bd \to \rho^0 K_2^{*0}$
and $B^+ \to \rho^0 K_2^{*+}$ decays in order to test the SM and this
type of NP.

\section{Conclusions}

We have seen that penguin annihilation and rescattering are two
possible SM solutions to the polarization puzzles in  $B\to\phi K^*$
and $B^{0(+)}\to \rho^{0(+)} K^{*0}$ ($\btos$) decays. Both potential
explanations also  predict a large $\fTfL$ ratio in certain $B \to V_1
V_2$ ($\btod$) processes. We have stressed that if it is found that
the transverse polarizations are large in $\bd \to \rho^0\rho^0$,  it
may be possible to test these explanations by comparing
$\bd \to \rho^0\rho^0$ with $B^+ \to K^{*0} \rho^+$ and see if flavor
SU(3) is respected. We have examined other $\btod$ decays related by
U-spin to certain $\btos$ decays.  Two promising pairs are: (i) $B^+
\to K^{*0} \rho^+$ ($\btos$) and $B^+ \to \Kbar^{*0} K^{*+}$ ($\btod$)
and (ii) $\bs \to K^{*0} \Kbar^{*0}$ ($\btos$) and $\bd \to \Kbar^{*0}
K^{*0}$ ($\btod$). A large $\fTfL$ is predicted by penguin
annihilation or rescattering in these decays. We have also mentioned
that it is possible to distinguish penguin annihilation from
rescattering by performing a time-dependent angular analysis of $\bd
\to \Kbar^{*0} K^{*0}$. This is difficult experimentally, but it may
be possible at a future machine.
 
Furthermore, we have analyzed the predictions of the SM (naive and
extended) and NP in $B \to V T$ decays. The SM naively reproduces the
polarization measurements in $B\to\phi K_2^*$ and predicts in general
a small $\fTfL$ ratio. The polarization predictions of both penguin
annihilation and rescattering are not certain. That is, the
predictions depend on a new set of parameters for each final state. It
is therefore possible that both explanations agree with the $\fTfL$
measurements in $B\to\phi K^*$ and $B\to\phi K_2^*$. Finally,  in
Ref.~\cite{NP} it was found that only two new-physics operators can
account for the discrepancies in both the $\pi K$ data and the
$\phi(\rho) K^*$ polarization measurements. We have mentioned that the
prediction of these operators can account for the small $\fTfL$ ratio
observed in $B\to\phi K_2^*$. This prediction can be tested by
explicit computations of the $B \to T$ form factors.

\bigskip 
\begin{acknowledgments}
I am grateful to A.~Datta, Y.~Gao, A.~V.~Gritsan, D.~London and
M.~Nagashima for their collaboration on this work. I also thank the
FPCP08 organizers for this wonderful conference.
\end{acknowledgments}

\bigskip 

\end{document}